\documentclass[genTeX]{nrc1b}

\usepackage{amssymb}
\newcommand{\vek}[1]{\mathrm{\bf #1}}

\volyear{83}{2005}
\journal{Can. J. Phys.}
\received{June 5, 2005}
\accepted{(date)}
\begin{document}

\title{Superheavy dark matter and ultrahigh energy cosmic rays}

\author[R. Dick]{R. Dick}
\address{Department of Physics and Engineering Physics,
        University of Saskatchewan, 116 Science Place, Saskatoon, 
        SK S7N 5E2, Canada. \email{rainer.dick@usask.ca}}
\author[K.M. Hopp]{K.M. Hopp}
\author[K.E. Wunderle]{K.E. Wunderle}

\shortauthor{Dick, Hopp and Wunderle}

\maketitle

\begin{abstract}
The phase of inflationary expansion in the early universe produces
superheavy relics in a mass window between $10^{12}\,$GeV and 
$10^{14}\,$GeV. Decay or annihilation of these superheavy relics can 
explain the observed ultrahigh energy cosmic rays beyond the 
Greisen-Zatsepin-Kuzmin cutoff. We emphasize that the pattern of 
cosmic ray arrival directions seen by the Pierre Auger observatory will 
decide between the different proposals for the origin of ultrahigh 
energy cosmic rays.
\end{abstract}

\begin{resume}
French version of abstract (supplied by CJP if necessary)

   \traduit
\end{resume}

\def\tablefootnote#1{
\hbox to \textwidth{\hss\vbox{\hsize\captionwidth\footnotesize#1}\hss}} 

\section{Introduction}\label{sec:intro}

The nature and origin of the dark matter in the universe and the origin
of ultrahigh energy cosmic rays (UHECR) are certainly two of the most 
interesting problems of astroparticle physics, and maybe even of physics, 
at the beginning of the 21st century. 

The three puzzles related to the origin of cosmic rays with 
energies beyond the Greisen-Zatsepin-Kuzmin cutoff, 
$E>E_{GZK}\simeq 4\times 10^{19}\,$eV, are:\\
{\bf 1.} Scattering off cosmic microwave
background photons limits the penetration depths of charged
particles at these energies to distances $<100\,$Mpc
\cite{GZK,stecker,new};\\
{\bf 2.} the distribution of arrival directions of UHECRs
does not seem to favor any known astrophysical sources within
the GZK cutoff length;\\
{\bf 3.} it seems extremely difficult to devise sufficiently efficient 
astrophysical acceleration mechanisms which could accelerate particles 
to energies $E>E_{GKZ}$, and at the same time overcome collisional
and radiation losses.

The so called top-down models of UHECRs combine both problems by proposing that
ultrahigh energy cosmic rays arise in the decay \cite{Hill,BKV,KR,BS,DFK} or
collisional annihilation \cite{BDK,DBK,DHW} of superheavy dark matter
particles. We will use both the acronym SHDM and 
{\sc wimpzilla}\footnote{The word {\sc wimpzilla} was coined by Kolb, 
Chung and Riotto in their investigations of possible origins of 
superheavy dark matter particles \cite{rocky}.} 
for superheavy dark matter particles.

SHDM decay can arise through direct decay of relic particles
(``{\sc wimpzilla}" decay) or through the annihilation of superheavy relic
bound states (``{\sc wimpzillium}" decay), but the underlying decay mechanism
plays no role for the predicted pattern of UHECR arrival directions.
This pattern differs strongly from the anisotropy pattern predicted by 
collisional {\sc wimpzilla} annihilation. The reason for the
difference is that the anisotropy pattern predicted by decay models
is proportional to the {\sc wimpzilla} or {\sc wimpzillium} density
$n_X(\vek{r})$, and is dominated by the smooth background halo. 
Collisional annihilation, on the other hand, predicts that the anisotropy 
pattern should be proportional to $n_X^2(\vek{r})$, and is
constrained by unitarity limits on annihilation cross sections. 
Therefore collisional annihilation can only work in dense cores of dark 
matter substructure in the galactic halo. These dense cores were denoted as
{\sc wimpzilla} stars \cite{DHW}. 
As a consequence, the collisional annihilation scenario predicts a pointlike
source distribution with increasing density towards the galactic center. 

All modern acceleration (``bottom-up'')
models for UHECR origin assume powerful extragalactic sources, and therefore
predict that the pattern of observed arrival directions should not correlate
with the galactic halo. The different patterns of UHECR arrival patterns
predicted by all the different source models imply that a dedicated 
large statistics experiment like the Pierre Auger observatory can easily
identify the correct model from its anisotropy signal.

Sec. \ref{sec:shdm} explains the origin of a mass window for the 
generation of relic superheavy dark matter particles during inflation.
Sec. \ref{sec:flux} summarizes and updates the calculation of the 
ultrahigh energy cosmic ray flux from collisional {\sc wimpzilla} 
annihilation, and the origin of the unique pattern of arrival directions 
predicted by that model. In Sec. \ref{sec:pattern} we compare the 
different anisotropy signatures predicted by all contemporary proposals 
for the origin of ultrahigh energy cosmic rays, and Sec. \ref{sec:conc} 
contains our conclusions.

\section{Superheavy dark matter from inflation}\label{sec:shdm}

Particle creation as a consequence of non-adiabatic expansion was 
discovered already in the late 1950s and 1960s, see \cite{parker} and 
references there. However, the effect was found to be negligible in 
radiation or dust dominated epochs, and therefore this mechanism for
non-thermal particle creation was rediscovered and garnered much more 
interest only after the necessity for inflation was realized.
The effect is usually considered in terms of the Bogolubov transformation
between in and out vacua in an expanding universe
\cite{rocky,CKR,KT,GPRT,CCKR}. Particle creation during 
preheating after inflation can also arise as a consequence of a direct 
coupling between the inflaton and other matter fields \cite{lev1,lev2}.

Here we will expand on a simple discussion given in \cite{DHW} to see
how particle production in an inflationary universe can 
be understood by studying the evolution equations of weakly coupled 
scalar fields in the expanding universe. The reasoning outlined here is
not intended to compete in any way with the traditional operator methods 
to study particle production from spatial expansion, but it may provide a
complementary and helpful view.

If interactions with other matter fields can be neglected, a scalar field 
in a Friedmann--Robertson--Walker background with metric
\begin{equation}\label{eq:frw}
ds^2=-dt^2+a^2(t)\!\left(
\frac{dr^2}{1-kr^2}+r^2d\vartheta^2+r^2\sin^2\vartheta\,d\varphi^2\right)
=-dt^2+a^2(t)\!\left(
d\vek{x}^2+k\frac{(\vek{x}\cdot d\vek{x})^2}{1-k\vek{x}^2}\right)
\end{equation}
satisfies
\begin{equation}\label{eq:phimot}
\ddot{\phi}(\vek{x},t)+3\frac{\dot{a}(t)}{a(t)}\dot{\phi}(\vek{x},t)
-\frac{1}{a^2(t)}\!\left[\left(\delta^{ij}-k x^i x^j\right)
\partial_i\partial_j-3kx^i\partial_i\right]\phi(\vek{x},t)
+m^2\phi(\vek{x},t)=0,
\end{equation}
and the corresponding energy density per unit of comoving volume is
\begin{eqnarray}\label{eq:rho}
\varrho(\vek{x},t)&=&\sqrt{-g}\,T^{00}
\\ \nonumber
&=&\frac{a^3(t)}{2\sqrt{1-kr^2}}\!\left[
\dot{\phi}^2(\vek{x},t)+m^2\phi^2(\vek{x},t)
+\frac{1}{a^2(t)}\!\left(\delta^{ij}-k x^i x^j\right)
\partial_i\phi(\vek{x},t)\cdot\partial_j\phi(\vek{x},t)\right].
\end{eqnarray}
The violation of time translation invariance in (\ref{eq:frw})
implies violation of energy conservation, of course:
\begin{eqnarray}\label{eq:rhodot}
\dot{\varrho}(\vek{x},t)&-&a(t)\partial_i\!\left[\dot{\phi}(\vek{x},t)
\frac{\delta^{ij}-k x^i x^j}{\sqrt{1-kr^2}}\partial_j\phi(\vek{x},t)\right]
\\ \nonumber
&=&\frac{a^2(t)\dot{a}(t)}{2\sqrt{1-kr^2}}\!\left[
3\!\left(m^2\phi^2(\vek{x},t)-\dot{\phi}^2(\vek{x},t)\right)
+\frac{1}{a^2(t)}\!\left(\delta^{ij}-k x^i x^j\right)
\partial_i\phi(\vek{x},t)\cdot\partial_j\phi(\vek{x},t)\right].
\end{eqnarray}
{\it A priori} the sign of the expression on the right hand side
is indefinite, e.g. a rapidly evolving field of low mass and with
small spatial fluctuations loses energy during spatial expansion
as long as it evolves rapidly enough. However, Eq. (\ref{eq:phimot})
tells us that spatial fluctuations (which should never have been large
anyway for the Friedmann--Robertson--Walker {\it ansatz} to work) are 
soon negligible in the inflationary expanding universe 
$a(t)\propto\exp(Ht)$.
This leaves us with the simple equation
\begin{equation}\label{eq:phi2}
\ddot{\phi}(t)+3H\dot{\phi}(t)
+m^2\phi(t)\simeq 0.
\end{equation}
Approximately constant $H$ during inflation then yields for
the time evolution of the comoving energy density
\begin{eqnarray*}
\varrho(t)&\simeq&
\frac{1}{2}a^3(t)\left(\dot{\phi}^2(t)
+m^2\phi^2(t)\right)
\\
&\simeq&
A_+\exp\!\left(t\sqrt{9H^2
-4m^2}\right)+A_-\exp\!\left(-t\sqrt{9H^2
-4m^2}\right)+B.
\end{eqnarray*}
This implies a growing mode in the comoving energy density of weakly coupled
states with $m<1.5H\simeq 10^{14}\,$GeV. 
What is special about the superheavy particles is that their 
comoving energy density is conserved after inflation, because the 
behavior of massive $(m>t^{-1})$ weakly coupled states 
in the subsequent radiation and dust dominated backgrounds preserves 
their energy.
The asymptotic solution for weakly coupled massive states 
with $m>t^{-1}$ in such a background yields (with $\ell=3$ for dust,
$\ell=4$ for radiation)
\[
\phi(t)\propto
t^{-3/\ell}\cos\!\left(mt+\varphi\right),
\]
\[
\varrho(t)\propto a^3(t)t^{-6/\ell}\propto t^0,
\]
and this implies in particular that the comoving density of massive 
particles freezes out
at the end of inflation $(t\simeq 10^{-36}\,\mbox{s})$ if 
\[
m>t^{-1}\simeq 10^{12}\,\mbox{GeV}.
\]
These considerations indicate a mass window for direct gravitational production
of superheavy relic particles during inflation
\[
10^{12}\,\mbox{GeV}<m<10^{14}\,\mbox{GeV}.
\]

After inflation intrinsically unstable matter of mass $M$ will
decay on time scales $\tau\lesssim M_{Planck}^2/M^3$, where the upper
bound assumes that the particles couple to their decay products at least
with gravitational strength. This means that {\it intrinsically unstable}
superheavy particles should not be relic particles. This difficulty
has motivated the collisional annihilation scenario for ultrahigh energy
cosmic rays from superheavy dark matter \cite{BDK}, because the unitarity
limit on reaction cross sections
\begin{equation}\label{eq:sigma1}
\langle\sigma_A v\rangle\lesssim\frac{4\pi}{M^2v}
\end{equation}
implies that superheavy dark matter without direct decay channels
will still be around.

\section{Calculation of the flux}\label{sec:flux}

We consider decay or annihilation of superheavy
dark matter particles of mass $M_X\ge 10^{12}\,$GeV.
The spectral fluxes at our location $\vek{r}_\odot$
from decay or collisional annihilation of the
dark matter particles of density $n_X(\vek{r})$ are then
\begin{equation}\label{eq:jdec}
j_d(E)=\frac{d\mathcal{N}(E,M_X)}{dE}
\int d^3\vek{r}\,\frac{1}{4\pi|\vek{r}_{\odot}-\vek{r}|^2}
\frac{n_X(\vek{r})}{\tau_d}
\end{equation}
and
\begin{equation}\label{eq:jann}
j_a(E)=\frac{d\mathcal{N}(E,2M_X)}{dE}
\int d^3\vek{r}\,\frac{\nu}{16\pi|\vek{r}_{\odot}-\vek{r}|^2}
n_X(\vek{r})^2 \langle\sigma_A v\rangle,
\end{equation}
respectively. Here
$d\mathcal{N}(E,E_{in})$ is the number of particles
in the energy interval $[E,E+dE]$ emerging from a decay or annihilation
event of initial energy $E_{in}$. 
$d\mathcal{N}(E,E_{in})/dE$ is related to fragmentation 
functions via
\[
\frac{d\mathcal{N}(E,E_{in})}{dE}=\sum_i\frac{1}{\sigma_A}
\frac{d\sigma^{(i)}}{dE}
=\frac{1}{E_{in}}\sum_i F^{(i)}(x,E_{in}^2),
\]
where $x=E/E_{in}$ and $F^{(i)}(x,E_{in}^2)$ is the differential
number of particles of species $i$ generated in the prescribed 
$x$-range in a decay or annihilation event with $s=E_{in}^2$.
The factor $\nu$ in Eq. (\ref{eq:jann}) equals 4 if the $X$-particles
are Majorana particles, and 1 otherwise.

$N$-body simulations predict that usually about $f_{cl}\sim 5-10$\% of 
dark matter halos should exist in substructure. Therefore in the decay 
models the ultrahigh energy cosmic ray flux will be dominated by the
smooth background halo. Evaluating (\ref{eq:jdec}) e.g. for a
Navarro-Frenk-White halo \cite{NFW}
\[
n_X(r)=4n_X(r_s)r_s^3/r(r+r_s)^2
\]
yields
\begin{equation}\label{eq:jdecnfw}
j_d(E)=4\frac{d\mathcal{N}(E,M_X/2)}{dE}
\frac{n_X(r_s)}{\tau_d}\frac{r_s^3}{r_s^2-r_{\odot}^2}
\ln\!\left(\frac{r_s}{r_{\odot}}\right).
\end{equation}
The scale radius $r_s$ for the Milky Way is not well known,
but for the whole range of say $5\,\mathrm{kpc}\le r_s\le
50\,\mathrm{kpc}$ comparison between (\ref{eq:jdecnfw})
and the ultrahigh energy cosmic ray flux from AGASA \cite{agasa}
yields that decaying superheavy dark matter should only 
make a small contribution to the galactic dark matter
density. Of course, the problem is to explain a lifetime
$\tau_d\ge 10^{17}\,$s for unstable superheavy dark matter.

Eq. (\ref{eq:jann}) can also be evaluated analytically for
a Navarro-Frenk-White halo, but the unitarity limit for
s-wave annihilation
\begin{equation}\label{eq:sigma}
\langle\sigma_A v\rangle=\xi\times\frac{4\pi}{M_X^2v}
=\xi\times 4.40\times 10^{-43}\,\mbox{m}^3/\mbox{s}
\times \left(\frac{10^{12}\,\mathrm{GeV}}{M_X}\right)^2
\times \frac{100\,\mathrm{km/s}}{v},\quad
\xi\le 1,
\end{equation}
implies that any ultrahigh energy cosmic ray flux from
collisional annihilation in the background halo is
negligible. Therefore collisional {\sc wimpzilla} annihilation
can only make a noticeable contribution to the ultrahigh energy 
cosmic ray flux if it originates in dense cores of
dark matter subclumps of the galactic halo \cite{BDK}.
These dense cores were denoted as {\sc wimpzilla} stars
in \cite{DHW}, and a simplified estimate of the cosmic ray flux 
from {\sc wimpzilla} stars yields
\begin{equation}\label{eq:flux1}
j_a(E)\simeq\nu\frac{N_{cl}\overline{V}_{core}}{16\pi d^2}
\frac{d\mathcal{N}(E,2M_X)}{dE}
n_X^2 \langle\sigma_A v\rangle
\simeq\nu\frac{0.1f_{cl}M_{halo}}{16\pi d^2 M_X}
\frac{d\mathcal{N}(E,2M_X)}{dE}
n_X \langle\sigma_A v\rangle.
\end{equation}
Here $d^{-2}=\langle r^{-2}\rangle$ is a mean inverse distance squared
for our separation from galactic {\sc wimpzilla} stars. In our estimate
in \cite{DHW} we had used $d\simeq 10\,$kpc, which seems reasonable: 
e.g. our inverse mean distance squared to visible galactic substructures, 
the globular clusters, corresponds to $d=7.3\,$kpc.

We have parametrized the unknown annihilation cross section already in 
terms of the unitarity bound (\ref{eq:sigma}). Now we would like to
compare e.g. with the UHE cosmic ray spectrum in Ref. \cite{BDK}, 
Fig. 1 (note that the cross section stated there would only be required 
for collisional annihilation in the smooth background halo
and does not apply to annihilation in dark matter subclump cores).
For the comparison we parametrize the core density $n_X$ of 
{\sc wimpzilla} stars in terms of the solar density
\[
n_X=\eta\times\frac{\varrho_\odot}{10^{12}\,\mbox{GeV}}
=\eta\times 7.89\times 10^{17}\,\mbox{m}^{-3}.
\]
This yields a flux at $10^{11}\,$GeV of order
\begin{equation}\label{eq:e3j}
E^3 j(E)\Big|_{E=10^{11}\,\mathrm{GeV}}
=5.74\times 10^{25}\,\mathrm{eV}^2\,\mathrm{m}^{-2}\,
\mathrm{s}^{-1}\times\frac{f_{cl}}{0.05}\times
\frac{M_{halo}}{2\times 10^{12}M_\odot}\times
\left(\frac{7.3\,\mathrm{kpc}}{d}\right)^2\times\nu
\times\frac{\eta\xi}{10^{-3}}.
\end{equation}
This complies with the spectral fit for the flux per solid angle
in Ref. \cite{BDK}, Fig. 1, if $\eta\xi<1$, as would be expected
for consistency of the model:
If the {\sc wimpzillas} are not Majorana particles ($\nu=1$)
the required value for the fit is
$\eta\xi\simeq 9.8\times 10^{-4}$, and otherwise the required
value would be $\eta\xi\simeq 2.5\times 10^{-4}$.

%%%%%%%%%%%%%%%%%%%%%%%%%%%%%%%%%%%%%%%%%%%%%%%%%%%%%%%%%%%%%%%%%%%%%%%%%%%%%
\section{Expected anisotropy patterns beyond $20\,$EeV}\label{sec:pattern}

Fits of fragmentation functions to the ultrahigh energy cosmic ray spectrum
indicate that in the collisional annihilation scenario cosmic rays above
$20\,$EeV should be dominated by the fragmentation products of {\sc wimpzilla}
annihilation \cite{BDK}. Therefore the expected anisotropy patterns discussed
in this section only apply in this energy range.

The purpose of this section is to emphasize that the anisotropy pattern
observed by the Pierre Auger observatory above $20\,$EeV will provide a
crucial direct test of the different proposals for the origin of ultrahigh
energy cosmic rays: Bottom-up acceleration of charged paricles
in AGNs (see e.g. \cite{RB}) or gamma ray bursts \cite{GRB}, 
ultrahigh energy neutrinos travelling over cosmological
distances and creating Z-bursts \cite{zburst},
{\sc wimpzilla} or {\sc wimpzillium} decay
\cite{Hill,BKV,KR,BS,DFK}, or collisional annihilation in {\sc wimpzilla}
stars \cite{BDK,DHW}.\\

\begin{tabular}{|l|l|}
Origin of UHECRs & Expected anisotropy pattern\\
 & \\ \hline & \\
Collisional annihilation in {\sc wimpzilla} stars & About 1000 pointlike
 sources with increasing\\
 & density towards the galactic center.\\
 & No correlation with galactic SNRs.\\
 & \\ \hline & \\
{\sc wimpzilla} or {\sc wimpzillium} decay & 
Dominated by uniform increase towards galactic center. \\
 & \\ \hline & \\
Z-bursts & Approximately isotropic distribution with only weak\\
 & correlation to structure within $150\,$Mpc. \\
 & \\ \hline & \\
Bottom-up acceleration & Correlation with local superstructure 
within $150\,$Mpc. \\ & \\ \hline    
\end{tabular}
{}\\

Once fully operational, the Pierre Auger observatory should see more than
300 events per year with energies above $40\,$EeV. If the collisional
annihilation scenario is correct, the observatory should see a large number
of multiplets within its angular resolution, with increasing
density towards the galactic center.

\section{Conclusions}\label{sec:conc}

Collisional annihilation of superheavy dark matter particles as a source
model for ultrahigh energy cosmic rays has the advantage of naturally
explaining the absence of a GZK cutoff in the spectrum without correlation
to local AGNs, without the need of postulating an extremely powerful and 
efficient acceleration mechanism, and without the need to explain 
extremely long lifetimes of intrinsically unstable particles. It has
the disadvantage of having to postulate formation of a few relatively
dense cores of dark matter subhalos or subhalo remnants.

The two different top-down scenarios of decay or collisional annihilation 
imply qualitatively different anisotropy signals in cosmic ray arrival 
directions, and these also differ from the anisotropy signals predicted 
by the various bottom-up scenarios. What constitutes a statistically 
significant dataset to confirm or reject the different anisotropy 
patterns depends on the different patterns, of course. However, for the
collisional annihilation scenario the Pierre Auger observatory should see 
a large number of multiplets within its angular resolution already after 
about two years of full operation, because the UHECRs should arise from a 
limited number of pointlike sources in our galactic halo. Based on 
current particle physics extrapolations for chemical composition of 
cosmic rays to very high energies, bottom-up acceleration scenarios seem 
much more popular in the cosmic ray community. However, collisional 
{\sc wimpzilla} annihilation is not an overly speculative theory, and 
based on its expected anisotropy pattern it will be confirmed or rejected 
in a very short time frame. It would therefore seem prudent to await what 
the Pierre Auger observatory will tell us about patterns of UHE cosmic ray
arrival directions before jumping to any conclusions about the validity
of different models.

\end{document}